%% file: Stalin_Growth.tex
\documentclass[11pt]{article}
\usepackage[a4paper,margin=2.25cm]{geometry}
\usepackage[T1]{fontenc}
\usepackage[utf8]{inputenc}
\usepackage{newtxtext,newtxmath}
\usepackage{microtype}
\usepackage{setspace}
\usepackage{booktabs,longtable,array,tabularx}
\usepackage{graphicx}
\usepackage{float}
\usepackage{caption}
\usepackage[table]{xcolor}
\usepackage{amsmath}
\usepackage{natbib}
\usepackage{hyperref}
\usepackage{tikz}
\usetikzlibrary{arrows.meta,positioning,shapes.geometric}
\hypersetup{colorlinks=true,linkcolor=black,citecolor=blue!45!black,urlcolor=blue!45!black,pdftitle={Was Stalin Necessary? Counterfactual Evidence on Soviet Growth},pdfauthor={Ricardo Alonzo Fernandez Salguero}}
\definecolor{accent}{HTML}{173A5E}
\definecolor{soft}{HTML}{F3F4F5}
\definecolor{rulegray}{HTML}{B5B5B5}
\begin{document}

\begin{center}
{\Huge\bfseries\color{accent} Was Stalin Necessary?}\par
\vspace{0.35em}
{\Large Counterfactual Evidence on Soviet Growth}\par
\vspace{1.1em}
{\large Ricardo Alonzo Fern\'andez Salguero}\par
\vspace{0.25em}
\href{https://orcid.org/0000-0002-4189-961X}{ORCID: 0000-0002-4189-961X}\par
\end{center}

\vspace{0.7em}
\begin{abstract}
This article evaluates whether the exceptionally rapid growth of the Russian and Soviet economy after 1928 requires a specifically Stalinist explanation. The empirical design treats the historical claim as a sequence of increasingly demanding counterfactual comparisons rather than as a binary contest between observed growth and one preferred alternative. Long-run output series from the Global Macro Database and the Maddison Project are combined with synthetic control, augmented synthetic control, synthetic difference-in-differences, interactive-factor models, low-rank matrix completion, robust prediction benchmarks, historical analogues, exact spatial placebos, fake treatment dates, recovery placebos, donor-pool perturbations, endpoint sensitivity, and independent Soviet national-account reconstructions. A central methodological change concerns the New Economic Policy alternative. Mechanical extrapolation of the 1922--1928 recovery rate is rejected because it confounds trend growth with the closure of a war and civil-war output gap. The revised NEP counterfactual is estimated as a bootstrap recovery-gap process whose trend is anchored in the pre-1914 record and whose persistence is estimated from the observed 1921--1928 closure of the Russian output gap. The Russian-core specifications exceed the median admissible frontier at every endpoint and exceed the primary 90th-percentile frontier from the mid-1930s onward. The broader Soviet territorial aggregate remains more conservative and does not exceed that frontier. Donor perturbations preserve a positive Russian-core premium, but the small effective placebo set, weak pre-treatment fit for several estimators, recovery-related false positives, source dependence, and hostile upper bounds prevent an inference of unique or absolute historical necessity. The evidence supports a conditional growth premium and a narrower proposition: the observed acceleration required an unusually strong resource-mobilisation mechanism, while the available data do not identify Stalin as the only institutionally possible route.
\end{abstract}


\vspace{0.8em}
\noindent\fcolorbox{rulegray}{soft}{\parbox{0.94\textwidth}{\textbf{How to cite.} Fern\'andez Salguero, Ricardo Alonzo. 2026. \emph{Was Stalin Necessary? Counterfactual Evidence on Soviet Growth}. Zenodo. \href{https://doi.org/10.5281/zenodo.21180776}{https://doi.org/10.5281/zenodo.21180776}.}}

\section{The historical question and the counterfactual object}

The claim that Stalin was necessary for Soviet development contains several propositions that must be separated before statistical evidence can be interpreted. One proposition is descriptive: output per person and aggregate production rose rapidly after 1928. A second is comparative: the observed trajectory exceeded the path that would have emerged under plausible non-Stalinist alternatives. A third is institutional: an exceptional mechanism of compulsory saving, agricultural extraction, sectoral reallocation, and heavy-industrial investment was required to obtain that speed of transformation. A fourth is personal and absolute: Stalin himself was the unique historically possible cause. Only the first three propositions can be approached with the available historical panel evidence, and the degree of identification falls sharply as the claim moves from comparative growth toward uniqueness. The article therefore treats necessity as a bounded counterfactual statement rather than as an assertion about every logically imaginable historical world.

The formal target is the cumulative post-1928 difference
\[
\Delta_{t}=\log Y^{\mathrm{obs}}_{t}-\log Y^{0}_{t},
\]
where $Y^{\mathrm{obs}}_{t}$ is the observed output or output-per-capita path and $Y^{0}_{t}$ is an unobserved non-Stalinist counterfactual. No single estimator can reveal $Y^{0}_{t}$ without auxiliary assumptions. Synthetic control requires a credible convex combination of donor histories; synthetic difference-in-differences requires that unit and time weights approximate the untreated latent structure; generalized synthetic control and matrix completion require a stable low-rank factor representation; predictive models require temporal stability but do not independently identify causality; historical analogues require institutional comparability; and NEP continuations require a model of how much of 1922--1928 was recoverable gap closure rather than sustainable trend growth. The evidentiary strategy consequently combines estimators with different failure modes, then subjects their common conclusion to adversarial boundaries.

The design follows the comparative logic of synthetic control developed by \citet{abadie2003,abadie2010,abadie2015}, its bias-corrected extension in augmented synthetic control \citep{benmichael2021}, the unit-and-time weighting approach of synthetic difference-in-differences \citep{arkhangelsky2021}, interactive-factor counterfactuals \citep{xu2017}, and low-rank matrix completion \citep{athey2021}. Exact and conformal inference is used as a complement to, rather than a replacement for, historical falsification \citep{chernozhukov2021}. These tools are applied to a case in which the treatment is not a single statute. The post-1928 regime combined the first Five-Year Plan, rapid industrial reallocation, collectivisation, restrictions on consumption, compulsory mobilisation, and a growing defence burden. The object estimated here is therefore the effect of the realised post-1928 regime package relative to specified alternatives, not a separable coefficient for the personality of Stalin.

The distinction between the Russian core and the broader Soviet aggregate is essential. Historical series labelled RUS and SUN are not interchangeable. They differ in territorial composition, population denominators, and the reconstruction of regional production. A result that is strong for the Russian core but weaker for the broader Soviet aggregate cannot be summarised as a homogeneous effect on an invariant national unit. The analysis consequently reports both objects and treats their divergence as substantive evidence about measurement and territorial composition rather than as a nuisance to be averaged away.

\begin{figure}[H]
\centering
\includegraphics[width=0.98\textwidth]{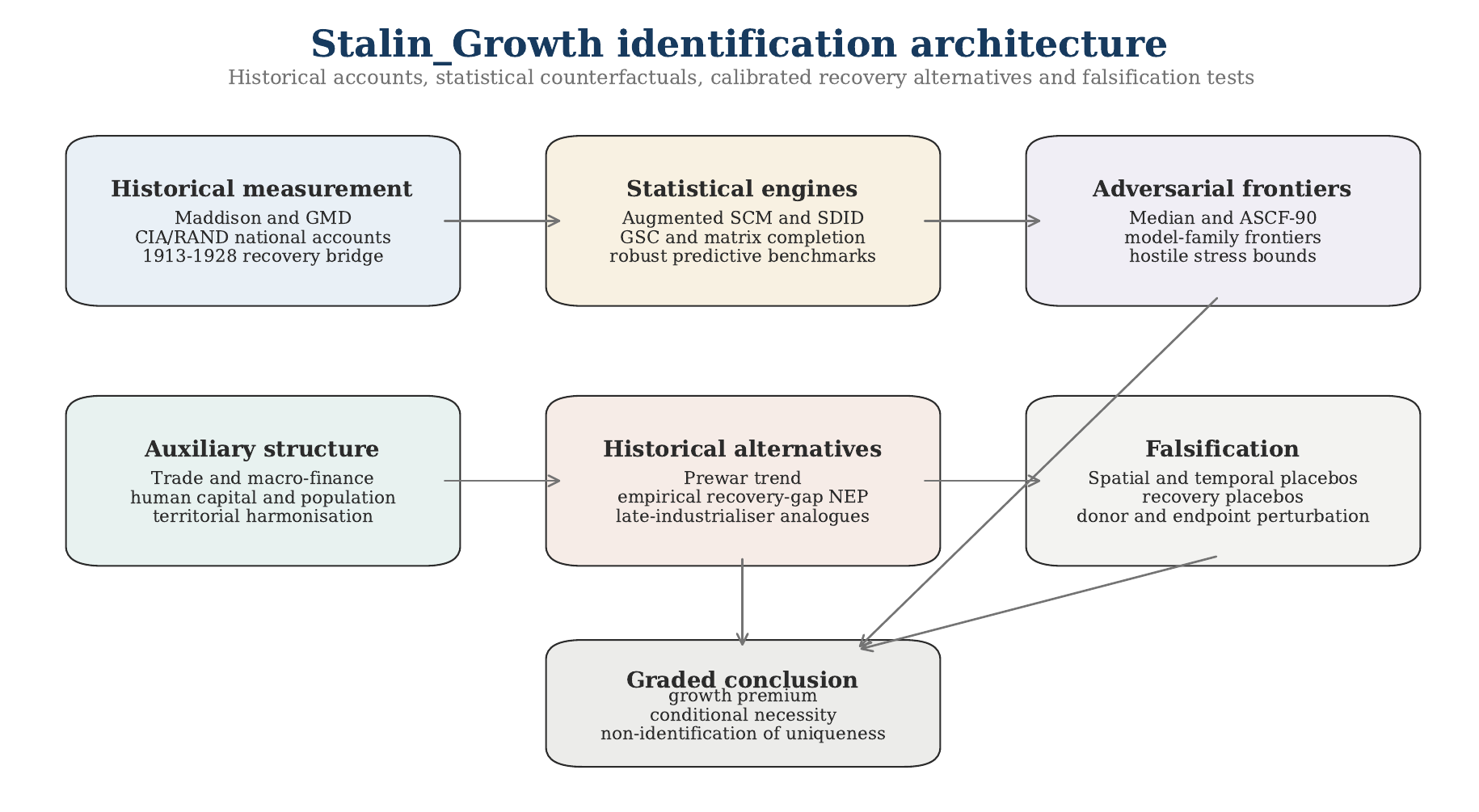}
\caption{Identification architecture. Historical measurements and auxiliary controls feed statistical counterfactual engines and historically calibrated alternatives. These paths are evaluated against admissible and hostile frontiers and subjected to falsification tests before the substantive claim is graded.}
\end{figure}

\input{tables/latex/table01_data_sources.tex}

The output series are drawn primarily from the Global Macro Database and the Maddison Project. Their agreement is informative but not equivalent to independent replication because historical reconstructions can share underlying sources. The national-account triangulation therefore also uses the declassified CIA/RAND reconstruction of Soviet GNP and the 1913--1928 bridge developed by \citet{markevich2011}. Maddison-style historical estimates are used as an alternative output source rather than as unquestioned observations of a latent true series \citep{bolt2024}. Macro-financial, trade, and educational datasets provide donor and shock diagnostics; the methodological precedent for using long-run macro-financial panels is represented by \citet{jorda2017}, while long-run educational controls follow the structure of \citet{barrolee2013}. The source architecture is deliberately redundant because the central historical uncertainty is not sampling error alone but the reconstruction of output, population, borders, and sectoral accounts.

\section{Data construction, estimators, and the recovery alternative}

Every post-treatment trajectory is normalised to $1928=100$. If the first observed post-origin log level is $y_{1929}$ and the first annual log growth is $g_{1929}$, the 1928 anchor is recovered as $y_{1928}=y_{1929}-g_{1929}$. Each observed and counterfactual path is then plotted as $100\exp(y_t-y_{1928})$. This normalisation exposes the timing and cumulative magnitude of divergence without pretending that different sources share the same level units. The principal endpoint is 1938 because it precedes the strongest effects of European war mobilisation and Soviet territorial change. Endpoints 1933, 1934, 1936, and 1940 identify the dynamics and test whether the conclusion depends on a favourable terminal date.

The primary statistical estimators are augmented synthetic control, synthetic difference-in-differences, generalized synthetic control, and matrix completion. Ordinary SCM remains informative but is not privileged when its pre-treatment fit is weak. Ridge, elastic net, Huber regression, partial least squares, and ensemble machine learning are retained as predictive benchmarks. Their role is to test whether a positive premium survives alternative forecasting rules, not to transform flexible prediction into causal identification. The model registry therefore assigns each component an inferential role before results are compared.

\input{tables/latex/table02_model_architecture.tex}

The original deterministic NEP lines are replaced by a recovery-gap model. The reason is historical and statistical. Average growth during 1922--1928 reflects the rebound from war, revolution, civil war, economic disintegration, and the 1921 trough. Treating that rate as a stationary NEP growth parameter generates an implausible exponential continuation. The revised construction first estimates a prewar log-linear trend from the Russian 1900--1913 series. It then defines the annual recovery gap as the distance between that prewar potential path and observed output per capita. The persistence of the gap is estimated from the sequence of gap ratios between 1921 and 1928. A residual bootstrap resamples uncertainty in the prewar trend and the annual gap-closure ratios. For each draw,
\[
\log Y^{\mathrm{NEP}}_{t}=\log Y^{\mathrm{potential}}_{t}-G_{1928}\rho^{t-1928},
\]
where $G_{1928}$ is the remaining log gap in 1928 and $\rho$ is the estimated annual persistence. This construction does not prove that the NEP would have continued unchanged. It creates a historically disciplined recovery benchmark whose uncertainty is explicit and whose median is not selected to match the observed post-1928 path.

\input{tables/latex/table05_nep_calibration.tex}

The bootstrap median places the NEP recovery alternative at an index of approximately 146 in 1940, with a 10--90 percent range of roughly 126--171. The observed Russian-core path reaches approximately 204. The difference is meaningful because it arises after the remaining recovery gap is modelled rather than after the exceptional 1922--1928 rate is copied into the future. The upper tail remains an adversarial historical alternative, but it no longer dominates the figure through an arbitrary slow-decay coefficient. Mechanical constant-NEP extrapolations are retained only in the stress archive and are excluded from the primary admissible frontier.

\begin{figure}[H]
\centering
\includegraphics[width=0.94\textwidth]{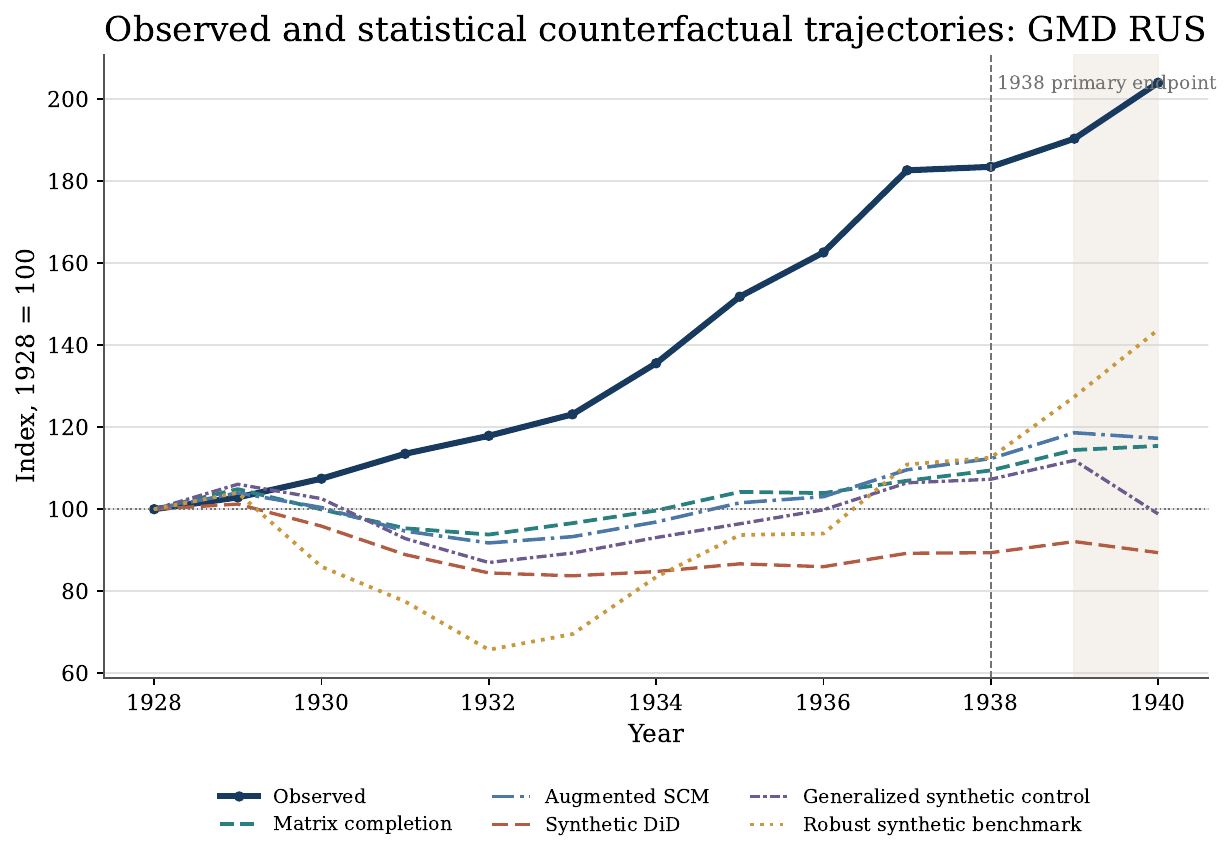}
\caption{Observed and statistical counterfactual trajectories for GMD RUS. The 1938 line marks the principal endpoint; 1939--1940 is treated as a mobilisation sensitivity.}
\end{figure}

The statistical paths diverge substantially in magnitude. Matrix completion combines the strongest pre-treatment fit with a positive post-1928 premium, while several unit-weight and factor estimators imply larger gaps but reproduce the pre-period less accurately. This heterogeneity is not hidden through a single model average. Instead, the article reports pre-treatment RMSE, pre-treatment correlation, conformal diagnostics, endpoint sensitivity, and model-family distributions. Large effects accompanied by weak pre-fit are evidence of model dependence, whereas a positive effect from a model with strong pre-fit receives greater weight.

\input{tables/latex/table04_model_diagnostics_gmd_rus.tex}

\begin{figure}[H]
\centering
\includegraphics[width=0.92\textwidth]{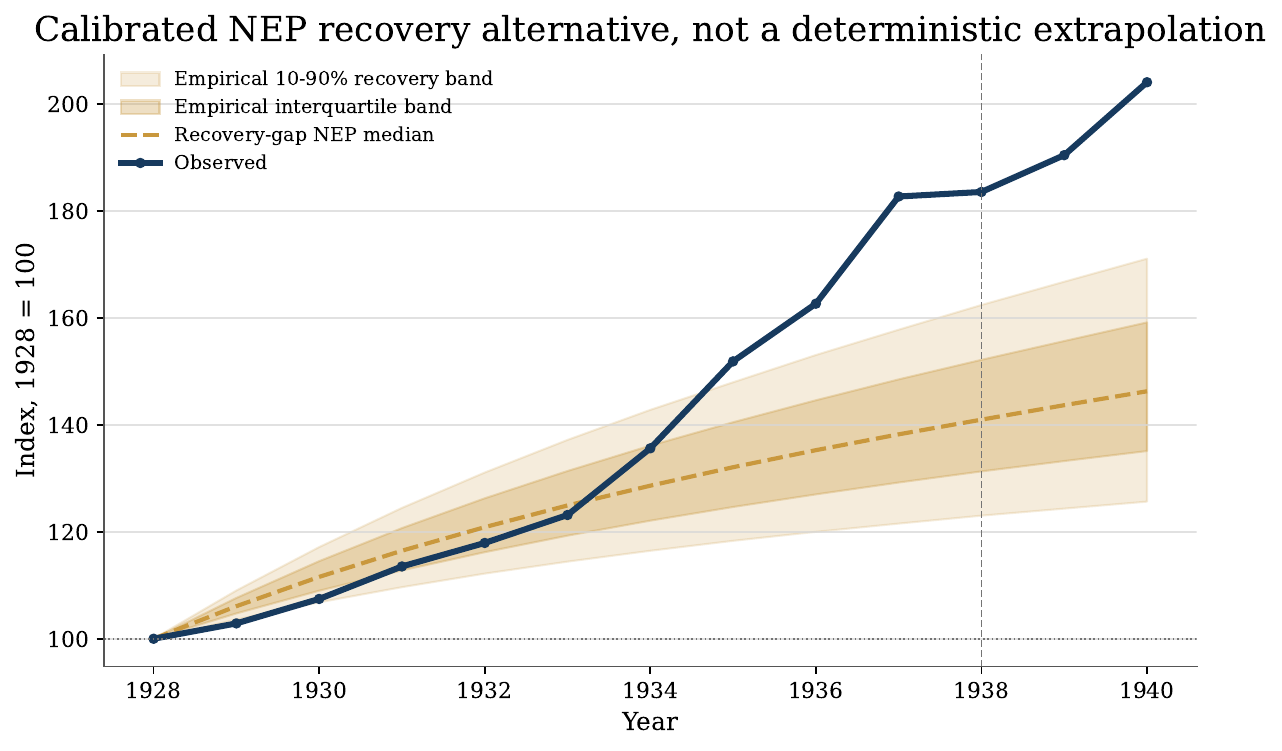}
\caption{Empirically calibrated recovery-gap NEP alternative. The bands reflect bootstrap uncertainty in the prewar trend and in the 1921--1928 gap-closing persistence.}
\end{figure}

Historical analogues are kept separate from statistical estimators. Japan and Turkey are realised paths of other state-led or late-industrialising economies, not identified substitutes for Soviet institutions. The year-specific donor percentile is even less interpretable as a coherent country because the country occupying the percentile can change from year to year. It is therefore labelled an adversarial envelope. Separating these objects prevents a descriptive historical trajectory from acquiring the inferential status of a synthetic-control estimate.

\begin{figure}[H]
\centering
\includegraphics[width=0.92\textwidth]{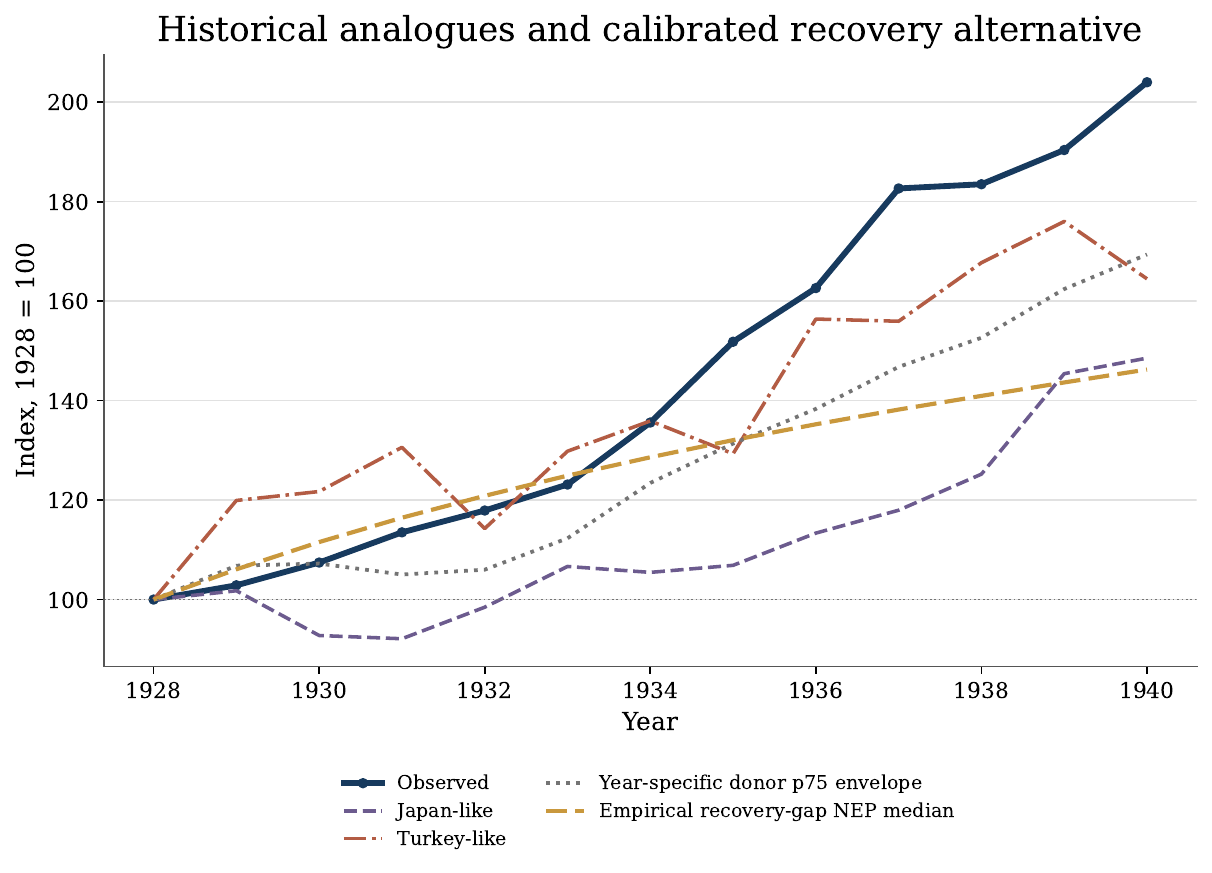}
\caption{Historical analogues and the calibrated recovery alternative. The donor percentile is an ex post annual envelope rather than a feasible institutional trajectory.}
\end{figure}

\section{Counterfactual evidence and endpoint stability}

The scorecard compares the observed cumulative path with the median admissible frontier, the 90th percentile of the admissible frontier, and a 90th-percentile stress frontier that also includes hostile upper bounds. Grade B denotes an observed path above the primary ASCF-90 frontier but below the stress frontier; grade C denotes an observed path above the primary median but below ASCF-90. This grading distinguishes a robust relative premium from a claim that defeats every hostile alternative.

\input{tables/latex/table03_main_scorecard.tex}

For GMD RUS, the observed path exceeds the median admissible frontier at every endpoint. The margin over ASCF-90 is negative in 1933 and 1934, turns positive in 1936, remains positive in 1938, and is still positive in 1940. Maddison RUS produces the same qualitative result and crosses the ASCF-90 frontier one endpoint earlier. Maddison SUN remains above the median frontier but below ASCF-90 throughout. The evidence is therefore not a uniform Soviet result. It is a stronger Russian-core premium combined with a more conservative broader territorial aggregate.

\begin{figure}[H]
\centering
\includegraphics[width=0.91\textwidth]{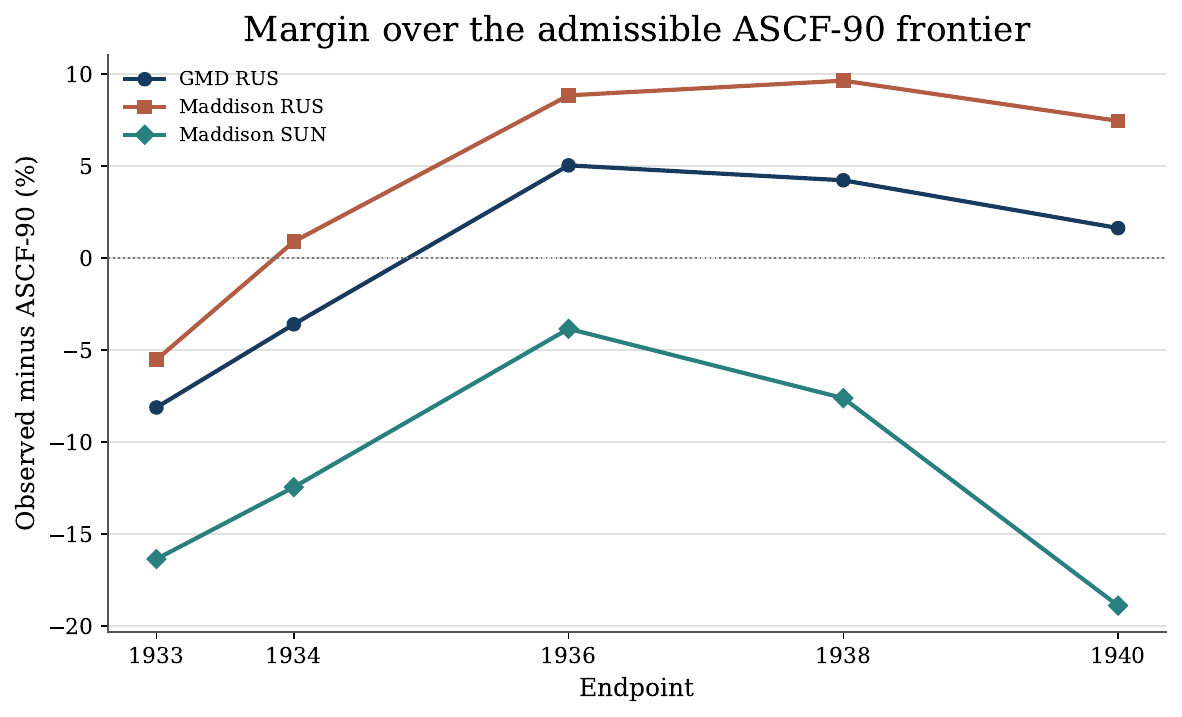}
\caption{Endpoint sensitivity of the margin over the primary ASCF-90 frontier. The Russian-core series cross the frontier before 1940; the broader SUN aggregate does not.}
\end{figure}

The endpoint pattern matters for causal interpretation. A result that appeared only in 1940 could reflect defence mobilisation, the European war shock, or territorial change. The positive Russian-core ASCF-90 margin in 1936 and 1938 shows that the comparative acceleration precedes the final endpoint. The shrinking GMD margin in 1940 also cautions against treating the terminal year as mechanically stronger evidence. The preferred substantive statement is that the Russian-core series moves above a demanding admissible frontier by the middle of the decade, not that every endpoint and territorial definition produces the same result.

\begin{figure}[H]
\centering
\includegraphics[width=0.85\textwidth]{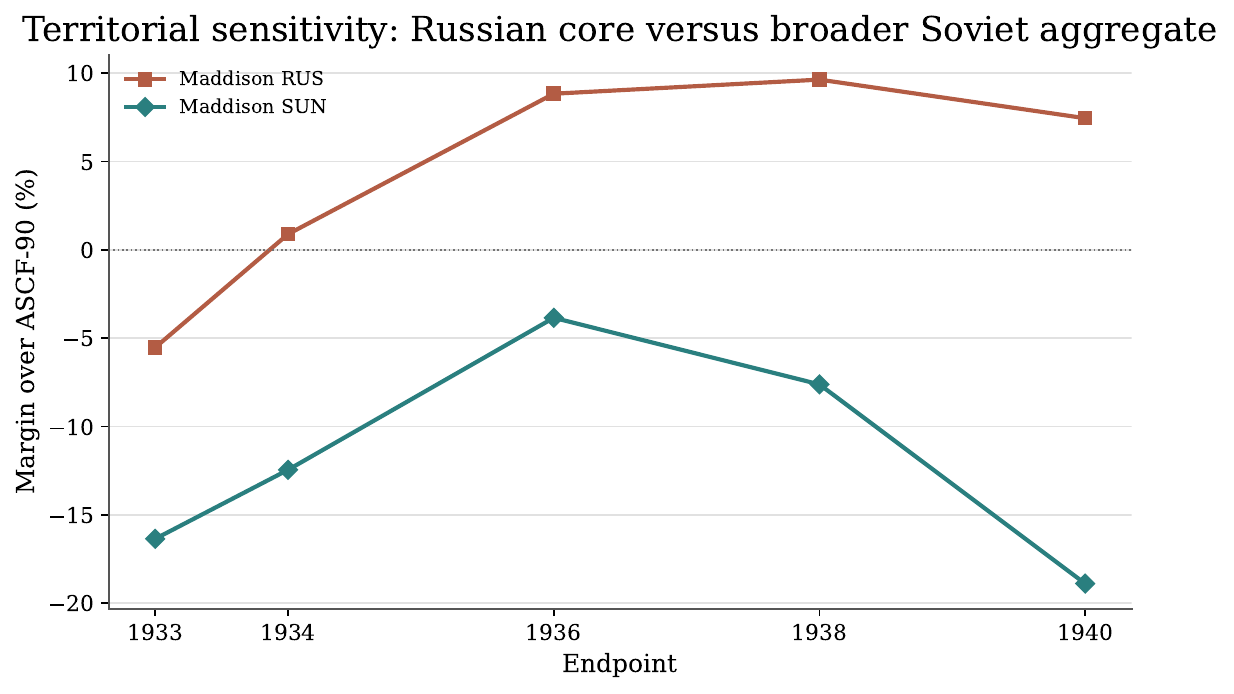}
\caption{Territorial sensitivity. The contrast between Maddison RUS and Maddison SUN shows that the historical unit of analysis materially affects the strength of the counterfactual claim.}
\end{figure}

The donor composition is economically interpretable but not innocuous. Italy, Sweden, Germany, and the United Kingdom carry the largest SCM weights in the principal 1940 specification. These weights do not imply that any one country is the historical alternative. They describe a convex combination selected to approximate the pre-treatment outcome trajectory. The concentration is moderate rather than complete, and the leave-one-donor-out exercises show that no single donor eliminates the positive premium. Nevertheless, a synthetic control that relies partly on fascist Italy and interwar Germany cannot be narrated as a clean contrast between Stalinism and liberal market development. The donor composition is an empirical prediction device whose institutional content requires separate interpretation.

\input{tables/latex/table06_top_donor_weights.tex}

\begin{figure}[H]
\centering
\includegraphics[width=0.84\textwidth]{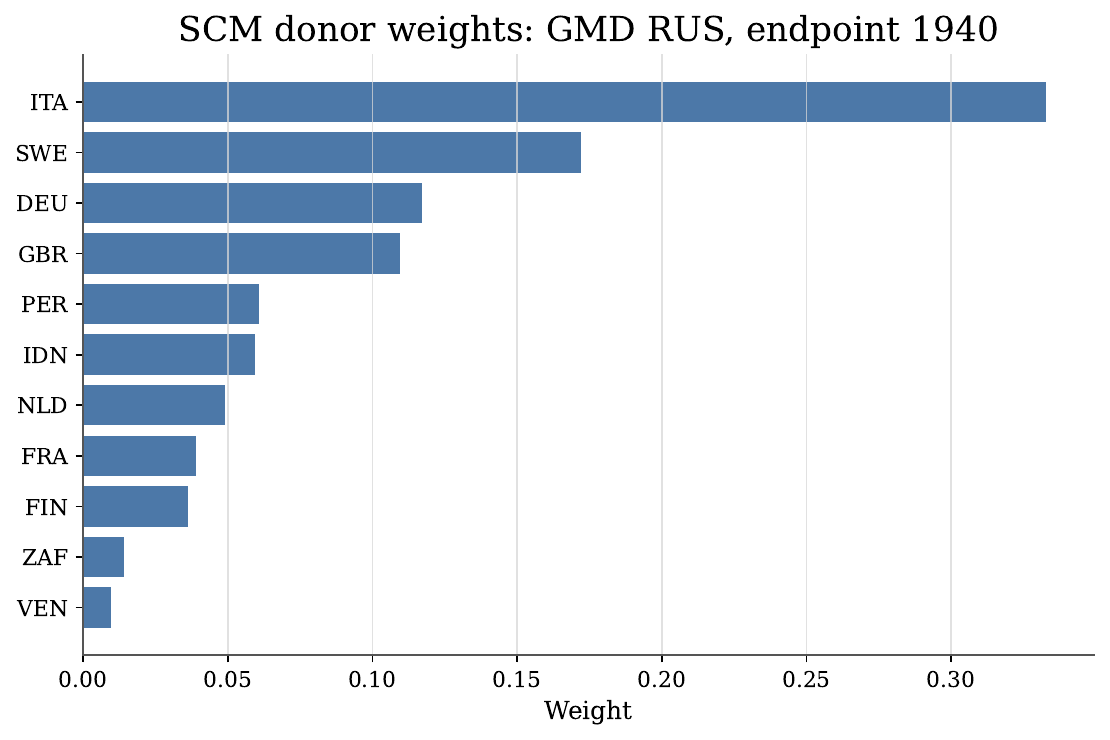}
\caption{Largest SCM donor weights for the GMD Russian-core specification through 1940.}
\end{figure}

The donor perturbation results are one of the stronger robustness findings. Removing each influential donor separately leaves a large positive cumulative premium. Randomly retaining 75 percent of the donor pool also produces a distribution centred near the all-donor result. This stability does not repair weak pre-fit, but it shows that the sign is not generated by one uniquely influential comparison country.

\input{tables/latex/table07_leave_one_out.tex}

\begin{figure}[H]
\centering
\includegraphics[width=0.84\textwidth]{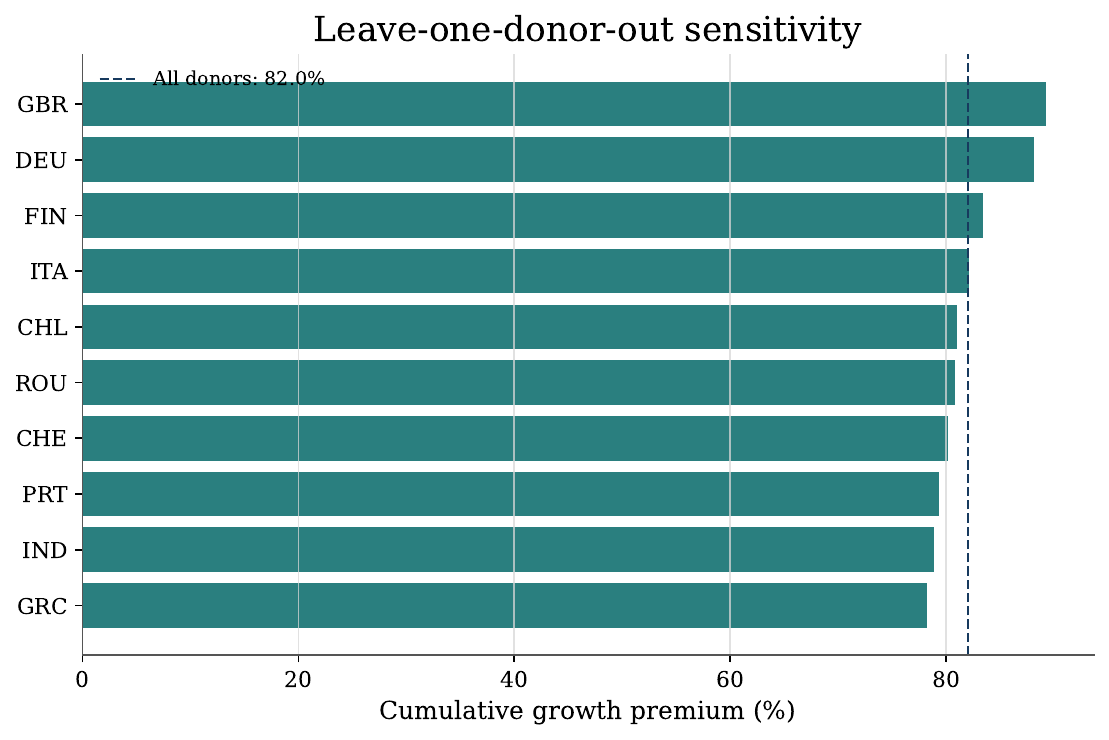}
\caption{Leave-one-donor-out sensitivity. The dashed line is the all-donor estimate.}
\end{figure}

\begin{figure}[H]
\centering
\includegraphics[width=0.84\textwidth]{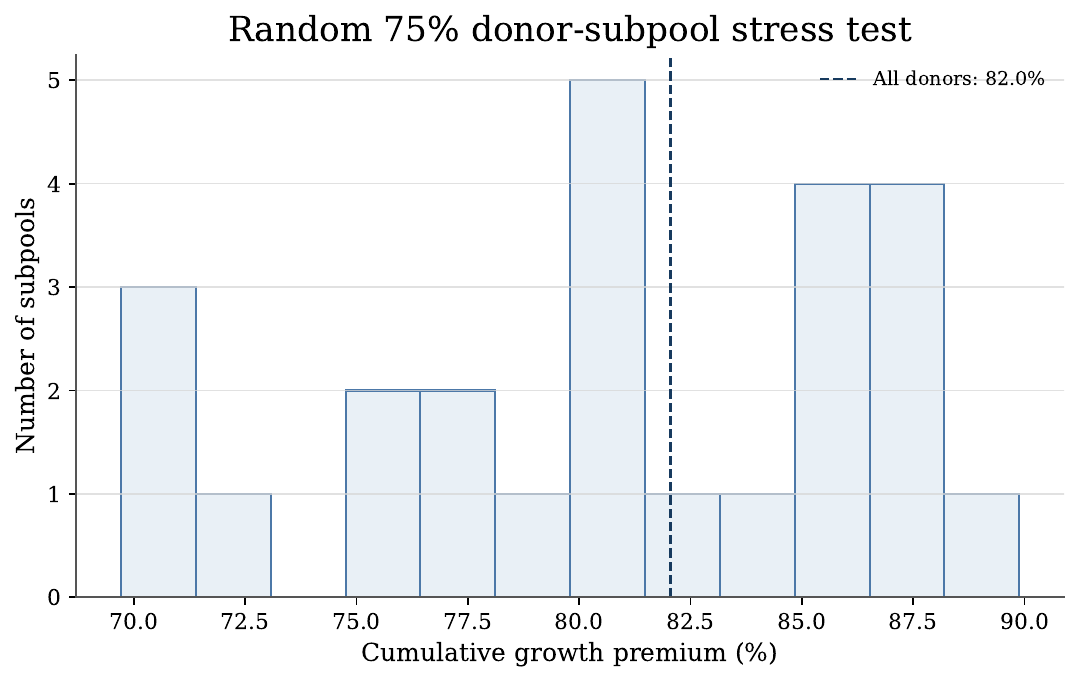}
\caption{Distribution of the growth premium under random 75 percent donor subpools.}
\end{figure}

The model-family distribution confirms that the positive sign is widespread but the magnitude is not point-identified. Low-rank methods, synthetic estimators, robust regressions, and historical scenarios occupy different parts of the effect distribution. The matrix-completion result is especially important because its pre-treatment fit is substantially better than the ordinary SCM, augmented SCM, synthetic DiD, and factor specifications. The result is therefore better summarised as a range of positive premiums under credible statistical specifications than as one precise percentage attributable to the regime.

\begin{figure}[H]
\centering
\includegraphics[width=0.88\textwidth]{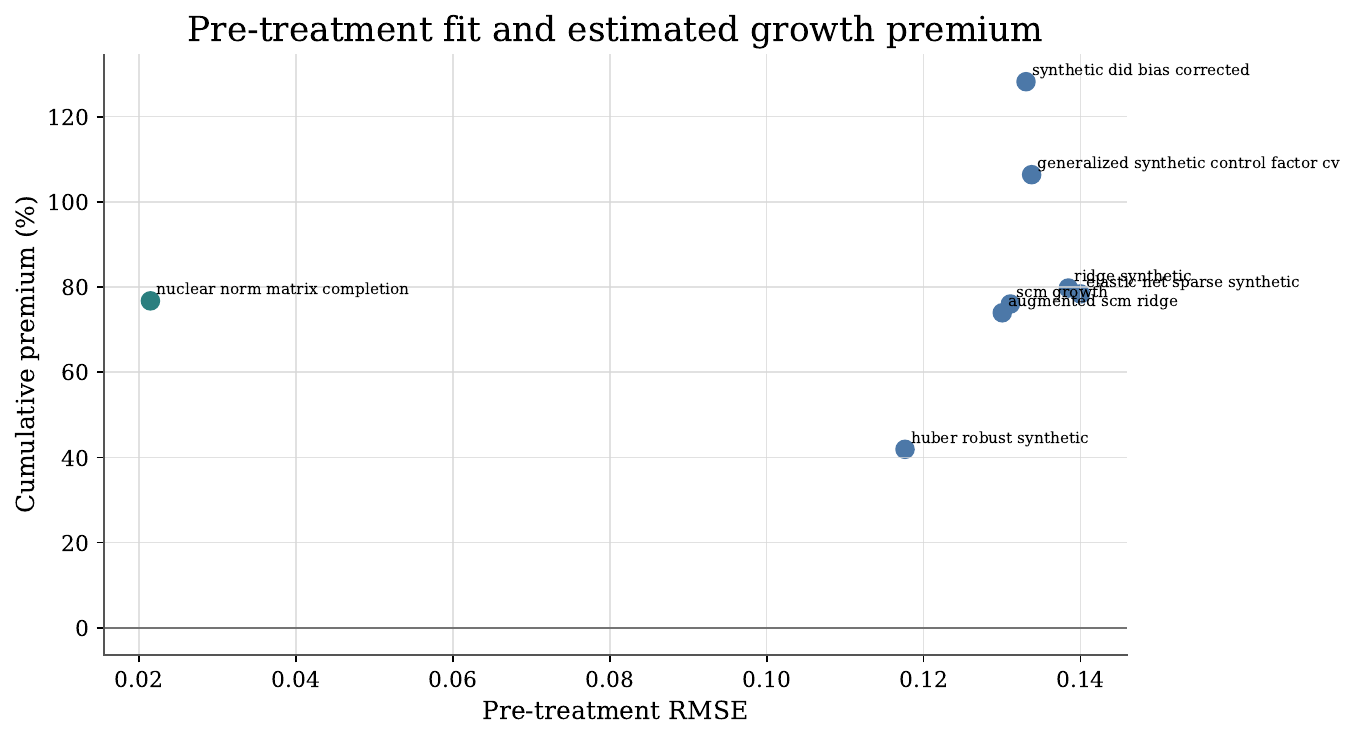}
\caption{Pre-treatment fit and estimated cumulative premium. Matrix completion occupies the low-error region; several larger estimates are associated with weaker pre-fit.}
\end{figure}

\begin{figure}[H]
\centering
\includegraphics[width=0.89\textwidth]{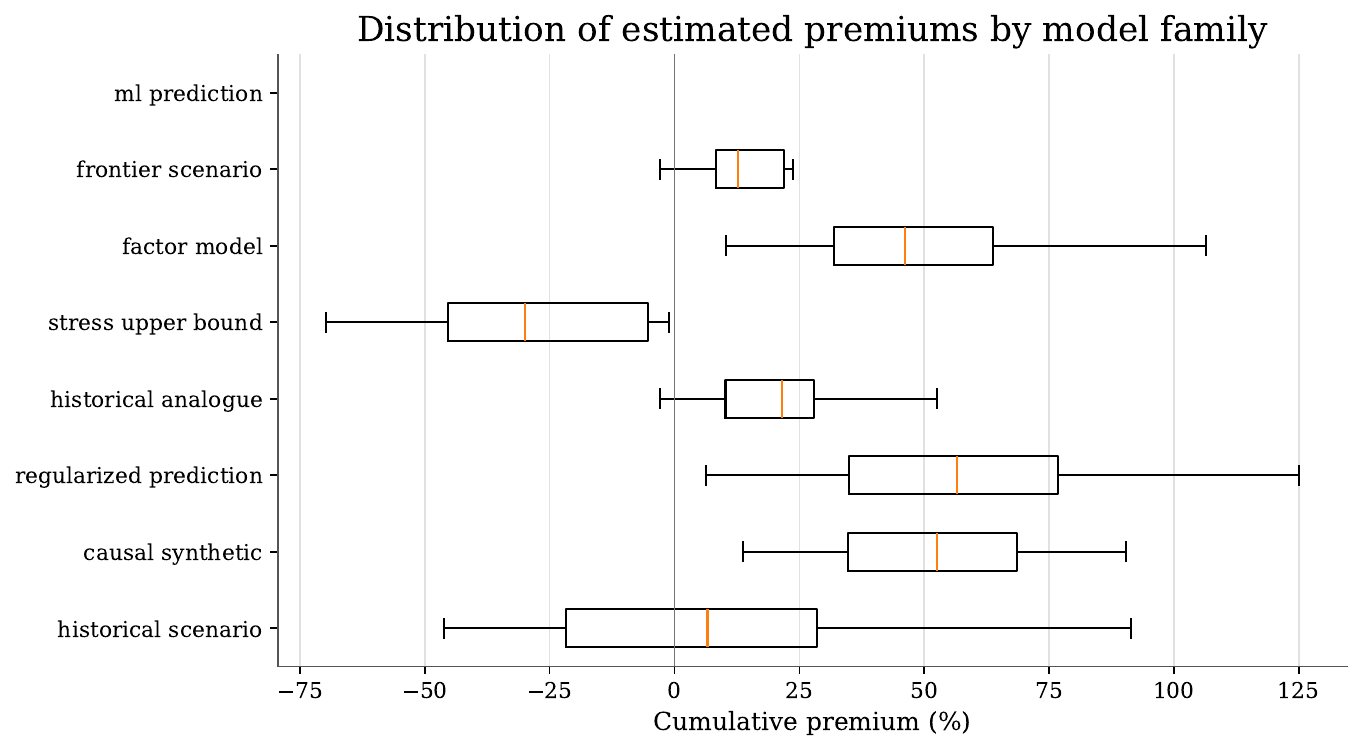}
\caption{Distribution of estimated cumulative premiums by model family.}
\end{figure}

\section{Falsification, national-account triangulation, and the limits of necessity}

Spatial placebos reassign the treatment to donor units and estimate effects using the same procedure. In the verified design there are eight effective placebo units for the key long-horizon specifications. The treated effect ranks first in several cases, producing an exact two-sided rank value of approximately $1/(8+1)=0.111$. This is suggestive but cannot satisfy conventional 10 or 5 percent thresholds because the placebo set is too small. The appropriate interpretation is rank dominance within a limited comparison set, not asymptotic statistical significance.

\input{tables/latex/table08_placebo_evidence.tex}

\begin{figure}[H]
\centering
\includegraphics[width=0.86\textwidth]{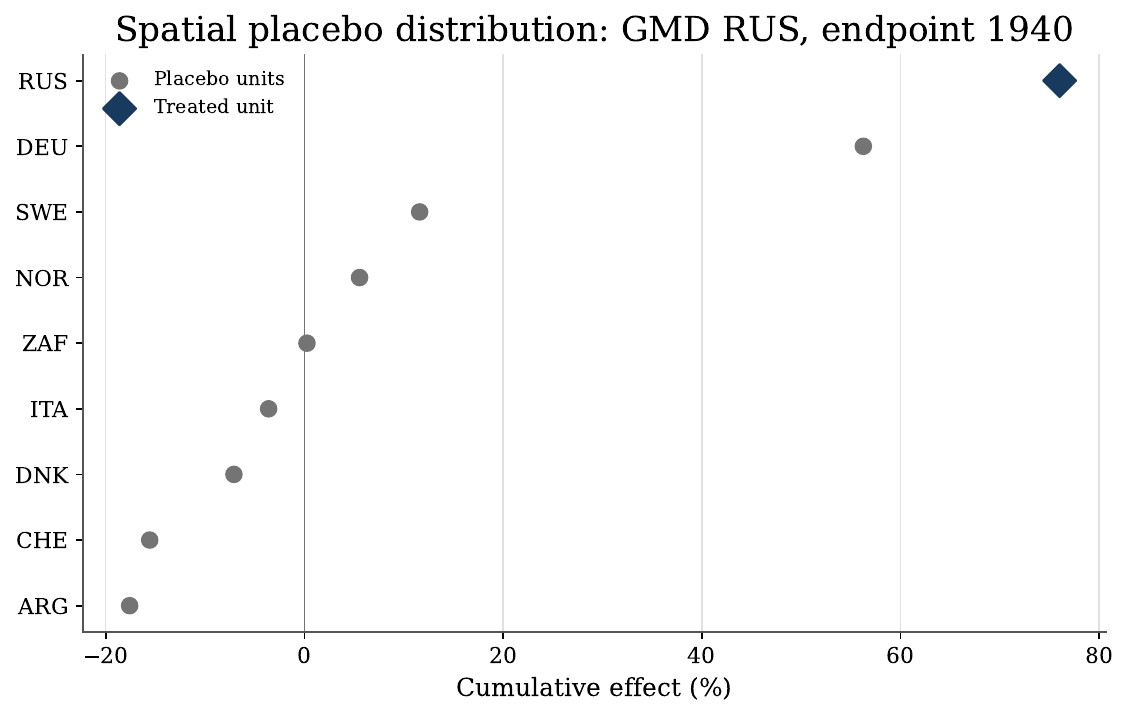}
\caption{Spatial-placebo distribution for GMD RUS through 1940. The treated unit is compared with the effective placebo set using the same estimator.}
\end{figure}

The post/pre RMSPE ratio provides a complementary diagnostic. It evaluates the post-treatment discrepancy relative to the estimator's own pre-treatment error. This matters because a large raw effect can be uninformative when the synthetic control already fits poorly before treatment. The ratio distribution should therefore be read together with the pre-RMSE table rather than as an independent significance test.

\begin{figure}[H]
\centering
\includegraphics[width=0.86\textwidth]{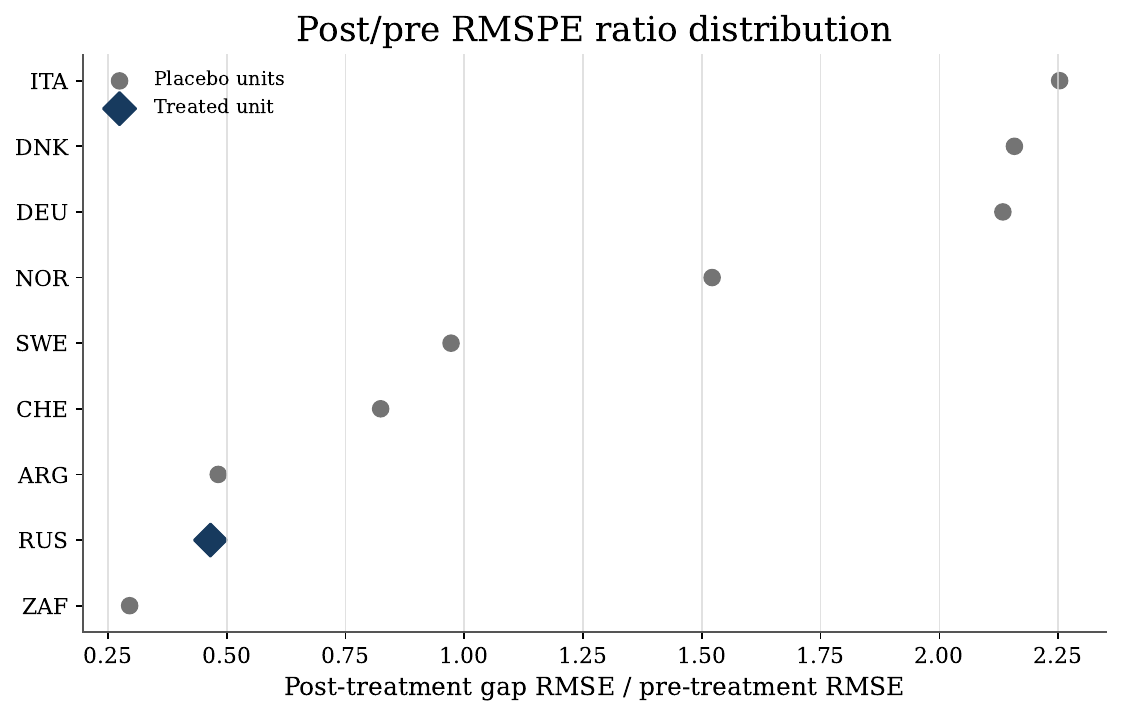}
\caption{Post/pre RMSPE ratios for the treated and placebo units.}
\end{figure}

Fake treatment origins expose a central threat: methods designed to detect post-1928 divergence can also detect the recovery from the 1921 trough. The pseudo-treatment assigned in 1921 generates a very large positive effect, while 1924 also produces a sizeable positive value. The effect becomes small near 1926. This pattern demonstrates why the NEP cannot be represented by an unbounded continuation of the early recovery rate. It also implies that the 1928 treatment effect partly inherits the historical problem of distinguishing structural transformation from the tail of post-conflict recovery.

\input{tables/latex/table09_time_placebos.tex}

\begin{figure}[H]
\centering
\includegraphics[width=0.86\textwidth]{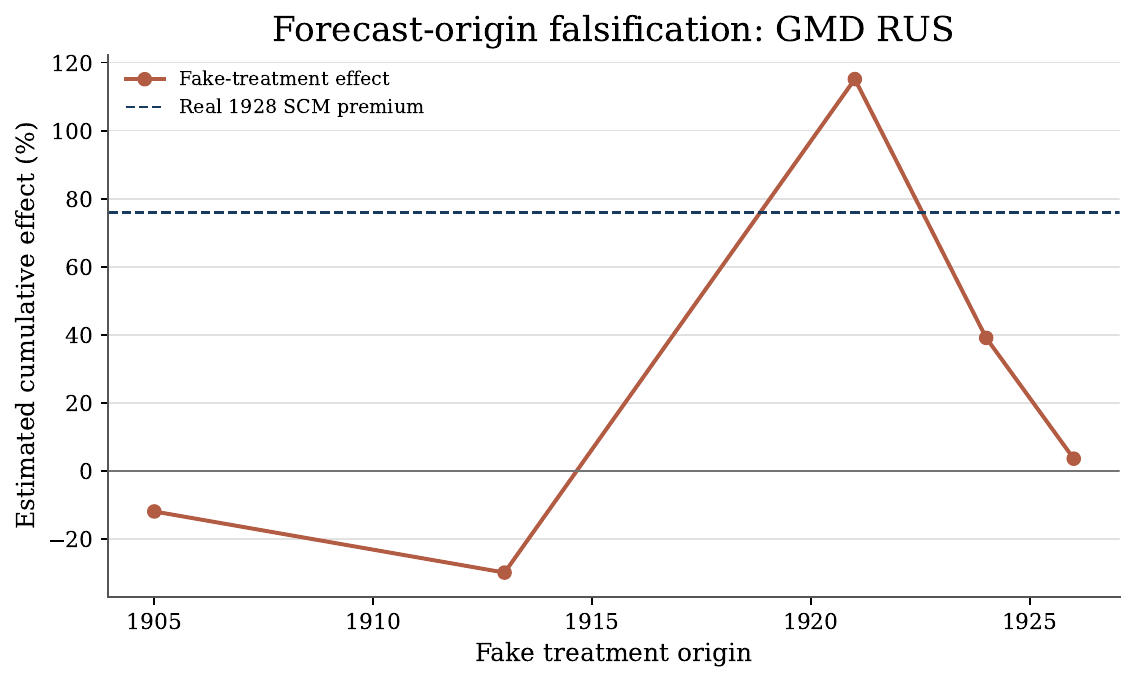}
\caption{Forecast-origin falsification. Large pseudo-effects after the 1921 trough reveal sensitivity to recovery dynamics.}
\end{figure}

Recovery placebos reinforce the same conclusion. Countries experiencing major preceding collapses can exhibit large pseudo-treatment effects during rebound. Greece, Belgium, South Africa, and an extreme Venezuelan episode demonstrate that an algorithm can mistake recovery for treatment when the pre-period contains a deep collapse and the post-period contains rapid normalisation. The Russian case cannot be discarded as an ordinary rebound because output in 1928 was already near the prewar level and the strongest divergence emerges later in the 1930s. It nevertheless requires explicit gap modelling, recovery-matched donors, and an interpretation that does not treat every post-collapse acceleration as policy causality.

\begin{figure}[H]
\centering
\includegraphics[width=0.85\textwidth]{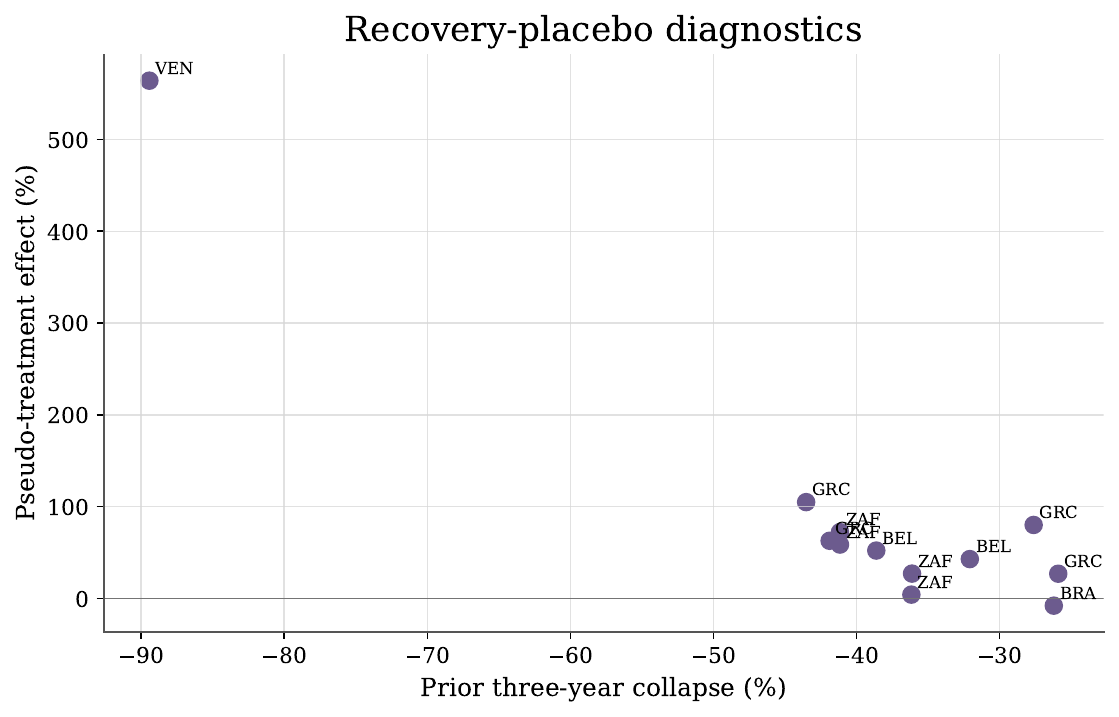}
\caption{Recovery-placebo diagnostics. Large pseudo-effects in other post-collapse episodes motivate the explicit recovery-gap alternative.}
\end{figure}

The independent national-account evidence is directionally consistent with high Soviet growth. CIA/RAND estimates imply average annual GNP growth of roughly 5.4--6.2 percent during 1928--1937 and 3.9--6.3 percent during 1937--1940. These estimates are not a direct validation of GMD or Maddison output per capita because they concern aggregate GNP and may share historical source material. Their value lies in showing that the high-growth pattern is not an artefact of one modern panel. The same accounts document a rising military share by 1940 and changes in the allocation between consumption, investment, and defence, making the terminal year substantively different from the earlier planning period. The broader literature on Soviet accounts, including \citet{bergson1961} and the structural comparison of \citet{cheremukhin2017}, likewise shows that the debate concerns the attainable alternative path and welfare cost, not the existence of industrial transformation itself.

\input{tables/latex/table10_cia_triangulation.tex}

\begin{figure}[H]
\centering
\includegraphics[width=0.82\textwidth]{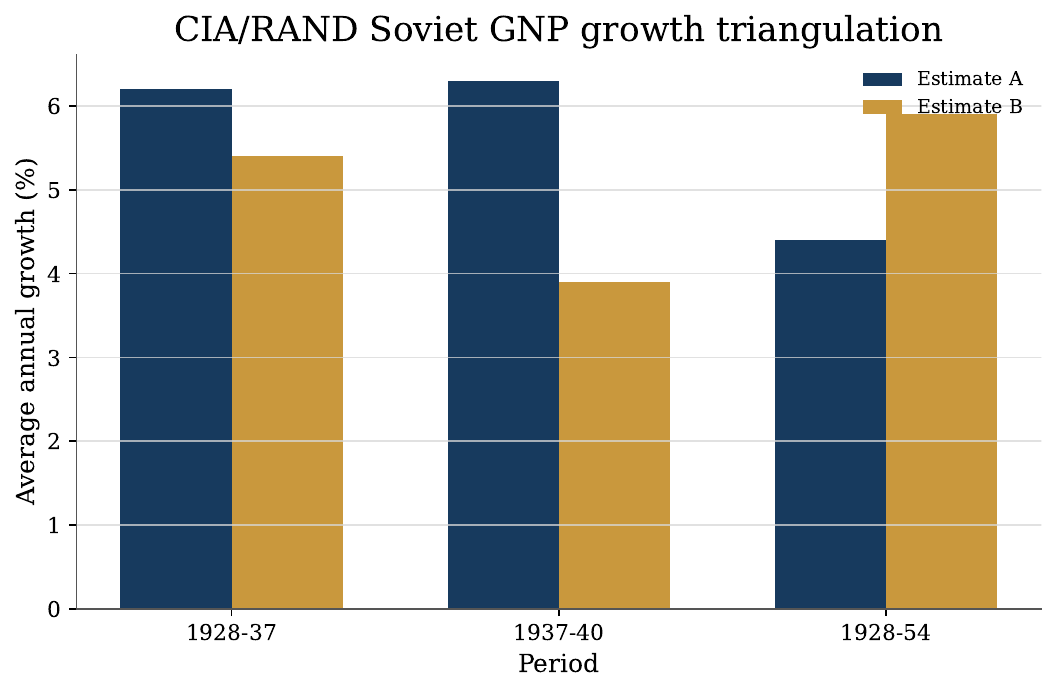}
\caption{CIA/RAND estimates of average annual Soviet GNP growth.}
\end{figure}

\begin{figure}[H]
\centering
\includegraphics[width=0.84\textwidth]{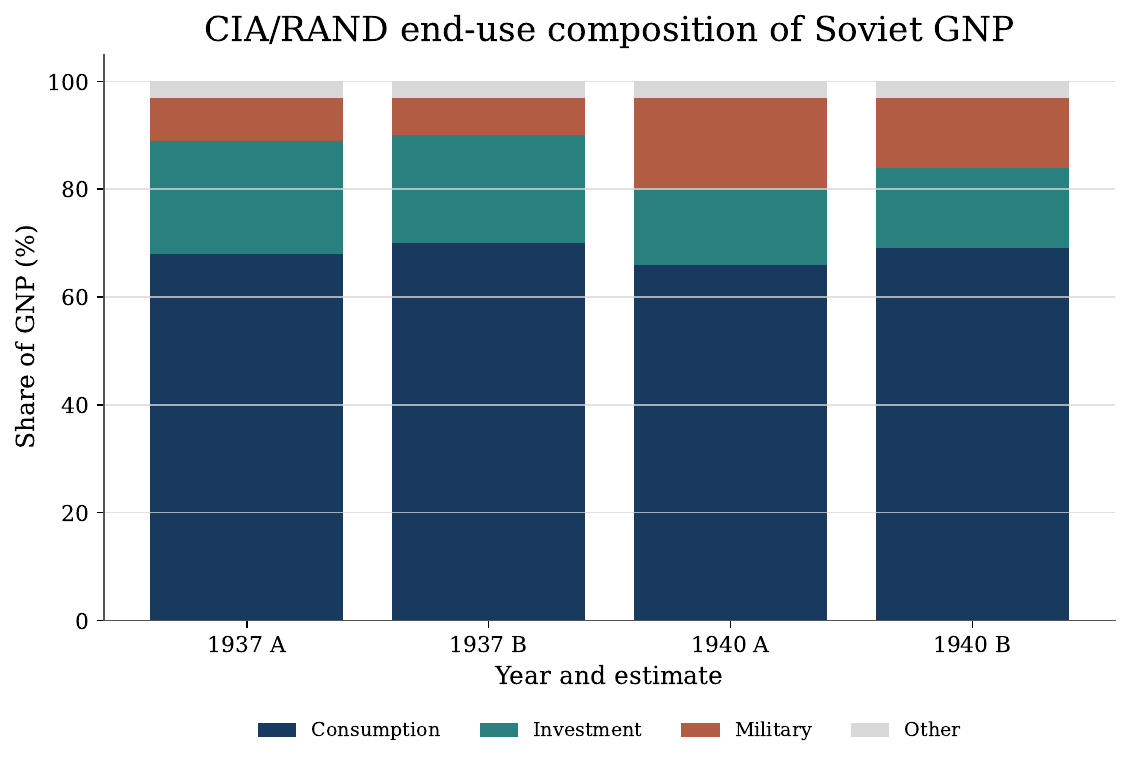}
\caption{CIA/RAND end-use composition. The rising military share clarifies why 1940 is treated as a mobilisation sensitivity rather than the sole primary endpoint.}
\end{figure}

The concept of necessity can now be stated precisely. A conditional necessity result would require the lower credible bound of the observed path to exceed the upper credible bound of every member of a prespecified admissible alternative set. The evidence does not satisfy that standard against all hostile bounds. The observed Russian-core path exceeds ordinary statistical counterfactuals, the calibrated median NEP recovery path, and the primary ASCF-90 frontier at several endpoints. It does not exceed every stress upper bound. Moreover, the admissible set cannot include every unobserved political arrangement that might have combined state capacity, external borrowing, market incentives, and coercion differently. The most defensible institutional conclusion is therefore narrower: a rapid and unusually forceful mobilisation of resources appears necessary to match the observed acceleration under the measured historical constraints, but the data do not uniquely identify Stalin's personal rule as the only mechanism capable of producing such mobilisation.

\begin{figure}[H]
\centering
\includegraphics[width=0.82\textwidth]{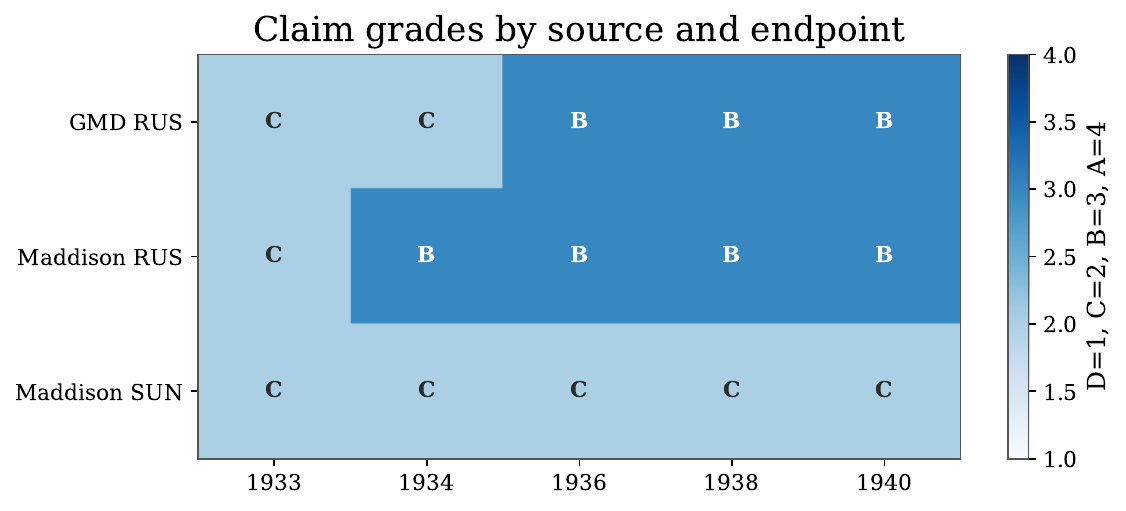}
\caption{Claim grades by source and endpoint. Grade B denotes an observed path above primary ASCF-90 but below the hostile stress frontier; grade C denotes superiority to the primary median only.}
\end{figure}

\input{tables/latex/table11_claim_scope.tex}

\section{Conclusion}

The post-1928 Russian-core growth path is difficult to reproduce with ordinary admissible counterfactuals. It exceeds the median frontier at all reported endpoints and moves above the primary 90th-percentile frontier by the middle of the 1930s in both GMD and Maddison reconstructions. The positive sign survives the removal of influential donors and random donor-subpool perturbations. Matrix completion produces a large premium while achieving substantially better pre-treatment fit than the classical SCM and several factor or weighting estimators. Independent Soviet national-account reconstructions are consistent with high aggregate growth and reveal the increasing role of defence by 1940.

The evidence is not equivalent to a proof that Stalin was uniquely necessary. The placebo set is small, several estimators have weak pre-treatment fit, effect magnitudes vary widely, recovery episodes generate false positives, and the broader SUN aggregate does not cross the primary ASCF-90 frontier. The NEP comparison is especially sensitive to the treatment of recovery. Once the exceptional 1922--1928 rate is replaced by a calibrated recovery-gap process, the median NEP path remains well below the observed Russian-core trajectory, but the upper band still represents meaningful uncertainty rather than a settled historical fact.

The resulting conclusion is conditional and cumulative. Soviet growth after 1928 displays a robust Russian-core premium relative to a broad set of statistical and historically disciplined alternatives. The acceleration is consistent with an institutional package capable of forcing saving, reallocating labour and capital, and prioritising heavy industry at exceptional speed. The analysis does not show that the same aggregate outcome could not have been reached by any other coercive or state-capacity configuration, nor does it convert output growth into a welfare judgement. The distinction between growth, institutional necessity, personal uniqueness, and welfare is not a rhetorical qualification; it is the identification boundary imposed by the historical data.

\end{document}

%% file: tables/latex/table01_data_sources.tex
\begin{longtable}{@{}p{0.28\textwidth}p{0.64\textwidth}@{}}
\caption{Historical data sources and their analytical roles.}\\
\toprule Source & Analytical role \\\midrule\endfirsthead
\toprule Source & Analytical role \\\midrule\endhead
Global Macro Database & Russian-core GDP and GDP-per-capita series; donor panels \\
Maddison Project Database 2023 & Alternative historical GDP-per-capita source and SUN/RUS robustness \\
CIA/RAND Soviet national accounts & GNP growth, sectoral origin and end-use triangulation \\
Markevich-Harrison appendix & 1913-1928 war, civil-war and NEP recovery bridge \\
COW/Federico-Tena/JST/Barro-Lee & Trade, macro-financial and human-capital diagnostics \\
\bottomrule\end{longtable}

%% file: tables/latex/table02_model_architecture.tex
\begin{longtable}{@{}p{0.17\textwidth}p{0.27\textwidth}p{0.48\textwidth}@{}}
\caption{Estimator families and inferential roles.}\\
\toprule Family & Estimator or block & Role \\\midrule\endfirsthead
\toprule Family & Estimator or block & Role \\\midrule\endhead
causal & Synthetic Control Method & Convex weights over donor growth paths; nonnegative weights sum to one. \\
causal & NNLS synthetic control & Nonnegative least squares, normalized donor weights. \\
causal & Augmented SCM & SCM plus ridge bias correction; reduces imperfect prefit bias. \\
causal & Synthetic Difference-in-Differences & Unit weights plus time/bias correction; compares post gaps after pre residual adjustment. \\
causal & Generalized Synthetic Control & Interactive fixed effects approximated by donor PCA/factor ridge. \\
causal & Matrix completion & Low-rank SVD/nuclear-norm-style panel imputation. \\
inference & Conformal trajectory bands & Residual-based conservative conformal intervals over post-treatment trajectory. \\
prediction & Ridge synthetic & Regularized donor regression. \\
prediction & Bayesian ridge & Posterior-shrinkage linear donor model. \\
prediction & Elastic net & Sparse regularized donor model. \\
prediction & Huber robust synthetic & Robust loss for outliers and crisis years. \\
prediction & PLS latent synthetic & Latent donor components optimized for treated pre-series. \\
ml & Random forest & Nonlinear benchmark with temporal-block validation warning. \\
ml & Extra trees & Nonlinear randomized-tree benchmark. \\
ml & Gradient boosting & Nonlinear additive-tree benchmark. \\
ml & kNN synthetic & Nearest-neighbor donor-year benchmark; flagged for overfit risk. \\
ml & Super learner & Ensemble of valid non-causal predictive models. \\
scenario & NEP slow reversion & High recovery gradually returns to long-run trend. \\
scenario & NEP fast reversion & Recovery gap closes quickly. \\
scenario & NEP capped & NEP growth capped by high historical donor percentile. \\
stress & NEP uncapped stress & Mechanical extrapolation of 1922-1928 recovery; not primary. \\
scenario & Tsarist mean trend & Prewar trend extrapolated. \\
scenario & Japan-like path & Late industrializer analogue. \\
scenario & Turkey-like path & Non-Soviet authoritarian modernization analogue. \\
scenario & Italy-like path & Corporate-authoritarian industrialization analogue. \\
frontier & Donor p75 frontier & High-growth admissible donor frontier. \\
stress & Donor p90 stress & Hostile upper bound. \\
stress & Population denominator stress & GDP per capita recomputed with alternative demographic assumptions. \\
triangulation & CIA/RAND sector and end-use check & Cross-checks GNP, industry, agriculture, consumption/investment/military use. \\
\bottomrule\end{longtable}

%% file: tables/latex/table05_nep_calibration.tex
\begin{table}[H]\centering
\caption{Bootstrap calibration of the empirical recovery-gap NEP alternative.}
\begin{tabular}{@{}lrrrp{0.33\textwidth}@{}}\toprule
Parameter & 5th pct. & Median & 95th pct. & Interpretation \\\midrule
prewar log-growth trend & 0.006 & 0.014 & 0.021 & annual log growth \\
annual recovery-gap persistence & 0.747 & 0.803 & 0.860 & AR(1)-style persistence \\
remaining log gap in 1928 & 0.123 & 0.236 & 0.350 & potential minus observed log output per capita \\
\bottomrule\end{tabular}\end{table}

%% file: tables/latex/table04_model_diagnostics_gmd_rus.tex
\begin{table}[H]\centering
\caption{Model diagnostics for GMD RUS through 1940.}
\begin{tabular}{@{}lrrrr@{}}\toprule
Estimator & Premium (\%) & Pre-RMSE & Pre-correlation & Outside band \\\midrule
nuclear norm matrix completion & 76.8 & 0.021 & 0.988 & Yes \\
huber robust synthetic & 41.9 & 0.118 & 0.551 & No \\
augmented scm ridge & 74.0 & 0.130 & 0.490 & No \\
scm growth & 76.0 & 0.131 & 0.496 & No \\
synthetic did bias corrected & 128.2 & 0.133 & 0.496 & No \\
generalized synthetic control factor cv & 106.4 & 0.134 & 0.302 & No \\
ridge synthetic & 79.8 & 0.138 & 0.271 & No \\
elastic net sparse synthetic & 78.4 & 0.140 & 0.102 & No \\
\bottomrule\end{tabular}\end{table}

%% file: tables/latex/table03_main_scorecard.tex
\begin{longtable}{@{}llrrrrc@{}}
\caption{Main counterfactual scorecard. Margins are percentage differences between observed cumulative growth and the corresponding frontier.}\\
\toprule Source & Treated & Endpoint & ASCF-50 & ASCF-90 & Stress-90 & Grade \\\midrule\endfirsthead
\toprule Source & Treated & Endpoint & ASCF-50 & ASCF-90 & Stress-90 & Grade \\\midrule\endhead
GMD & RUS & 1933 & 16.6 & -8.1 & -20.4 & C \\
GMD & RUS & 1934 & 27.0 & -3.6 & -19.6 & C \\
GMD & RUS & 1936 & 48.8 & 5.0 & -19.0 & B \\
GMD & RUS & 1938 & 63.0 & 4.2 & -23.2 & B \\
GMD & RUS & 1940 & 50.1 & 1.6 & -28.3 & B \\
Maddison & RUS & 1933 & 19.3 & -5.5 & -16.4 & C \\
Maddison & RUS & 1934 & 29.4 & 0.9 & -13.6 & B \\
Maddison & RUS & 1936 & 46.0 & 8.8 & -9.9 & B \\
Maddison & RUS & 1938 & 54.5 & 9.6 & -12.3 & B \\
Maddison & RUS & 1940 & 60.2 & 7.5 & -16.0 & B \\
Maddison & SUN & 1933 & 6.3 & -16.4 & -26.7 & C \\
Maddison & SUN & 1934 & 15.4 & -12.4 & -26.1 & C \\
Maddison & SUN & 1936 & 31.5 & -3.8 & -23.0 & C \\
Maddison & SUN & 1938 & 33.4 & -7.6 & -29.0 & C \\
Maddison & SUN & 1940 & 29.4 & -18.9 & -39.6 & C \\
\bottomrule\end{longtable}

%% file: tables/latex/table06_top_donor_weights.tex
\begin{table}[H]\centering
\caption{Largest SCM donor weights for GMD RUS, endpoint 1940.}
\begin{tabular}{@{}lrr@{}}\toprule Donor & Weight & Rank \\\midrule
ITA & 0.333 & 1 \\
SWE & 0.172 & 2 \\
DEU & 0.117 & 3 \\
GBR & 0.109 & 4 \\
PER & 0.061 & 5 \\
IDN & 0.059 & 6 \\
NLD & 0.049 & 7 \\
FRA & 0.039 & 8 \\
FIN & 0.036 & 9 \\
ZAF & 0.014 & 10 \\
\bottomrule\end{tabular}\end{table}

%% file: tables/latex/table07_leave_one_out.tex
\begin{table}[H]\centering
\caption{Most influential leave-one-donor-out specifications for GMD RUS.}
\begin{tabular}{@{}lrrr@{}}\toprule Excluded donor & Donors & Cumulative premium & Annualised premium \\\midrule
GRC & 27 & 78.2 & 4.9 \\
IND & 27 & 78.9 & 5.0 \\
PRT & 27 & 79.3 & 5.0 \\
CHE & 27 & 80.2 & 5.0 \\
ROU & 27 & 80.8 & 5.1 \\
CHL & 27 & 81.0 & 5.1 \\
ITA & 27 & 82.2 & 5.1 \\
FIN & 27 & 83.4 & 5.2 \\
DEU & 27 & 88.1 & 5.4 \\
GBR & 27 & 89.2 & 5.5 \\
\bottomrule\end{tabular}\end{table}

%% file: tables/latex/table08_placebo_evidence.tex
\begin{table}[H]\centering
\caption{Spatial placebo evidence for treated specifications with an effective placebo set.}
\begin{tabular}{@{}llrrrrr@{}}\toprule Source & Treated & Endpoint & Effect & Exact $p$ & Placebos & RMSPE ratio \\\midrule
Maddison & SUN & 1938 & 39.5 & 0.111 & 8 & 0.38 \\
Maddison & RUS & 1938 & 54.5 & 0.111 & 8 & 0.50 \\
GMD & RUS & 1938 & 65.3 & 0.111 & 8 & 0.47 \\
Maddison & SUN & 1940 & 35.1 & 0.222 & 8 & 0.36 \\
Maddison & RUS & 1940 & 65.1 & 0.111 & 8 & 0.49 \\
GMD & RUS & 1940 & 76.0 & 0.111 & 8 & 0.47 \\
\bottomrule\end{tabular}\end{table}

%% file: tables/latex/table09_time_placebos.tex
\begin{table}[H]\centering
\caption{Forecast-origin falsification tests for GMD RUS.}
\begin{tabular}{@{}rrrrrr@{}}\toprule Fake origin & Fake endpoint & Horizon & Pseudo-effect & Pre-RMSE & Post-gap RMSE \\\midrule
1905 & 1917 & 12 & -11.8 & 0.128 & 0.090 \\
1913 & 1925 & 12 & -29.8 & 0.122 & 0.179 \\
1921 & 1928 & 7 & 115.1 & 0.132 & 0.128 \\
1924 & 1928 & 4 & 39.2 & 0.133 & 0.108 \\
1926 & 1928 & 2 & 3.7 & 0.133 & 0.018 \\
\bottomrule\end{tabular}\end{table}

%% file: tables/latex/table10_cia_triangulation.tex
\begin{table}[H]\centering
\caption{CIA/RAND estimates of average annual Soviet GNP growth.}
\begin{tabular}{@{}lrrp{0.48\textwidth}@{}}\toprule Period & Estimate A & Estimate B & Interpretation \\\midrule
1928-37 & 6.2 & 5.4 & national-accounts triangulation for Soviet GNP growth \\
1937-40 & 6.3 & 3.9 & national-accounts triangulation for Soviet GNP growth \\
1928-54 & 4.4 & 5.9 & national-accounts triangulation for Soviet GNP growth \\
\bottomrule\end{tabular}\end{table}

%% file: tables/latex/table11_claim_scope.tex
\begin{longtable}{@{}p{0.31\textwidth}p{0.19\textwidth}p{0.42\textwidth}@{}}
\caption{Claims supported, rejected or left unidentified by the evidence.}\\
\toprule Claim & Assessment & Empirical basis \\\midrule\endfirsthead
\toprule Claim & Assessment & Empirical basis \\\midrule\endhead
Observed growth exceeds ordinary admissible counterfactuals & Supported & Strongest for GMD-RUS and Maddison-RUS from the mid-1930s onward. \\
Russian-core path exceeds the primary ASCF-90 frontier & Supported in several endpoints & Positive margins in 1936, 1938 and 1940 for GMD-RUS; from 1934 for Maddison-RUS. \\
Broader SUN territorial path exceeds ASCF-90 & Not supported & SUN beats the median frontier but not the primary ASCF-90 frontier. \\
The result survives donor perturbation & Supported & Leave-one-out and random-subpool tests preserve a positive premium in the principal RUS specification. \\
Spatial-placebo evidence is conventionally decisive & Not supported & Exact p-values are discrete and placebo support is limited in the verified v6 design. \\
Stalin as an individual was uniquely necessary & Not identified & The design supports a conditional growth premium, not metaphysical or personal necessity. \\
A comparable forced-mobilisation package may have been necessary & Partially supported & The conclusion depends on the admissible alternative set and on hostile NEP/donor upper bounds. \\
Aggregate growth establishes welfare superiority & Not supported & Growth is distinct from consumption, mortality, coercion and welfare. \\
\bottomrule\end{longtable}